\documentclass[a4paper]{raa_rb}       

\usepackage{graphicx,times}             
\usepackage{natbib}
\usepackage{amssymb,amsmath}
\bibpunct{(}{)}{;}{a}{}{,}
\usepackage{graphicx}
\usepackage{amsmath}
\usepackage{ulem}
\usepackage{booktabs}
\usepackage{hyperref}
\usepackage{xcolor}
\usepackage{makecell}

\begin{document}

\title{The Multi-parameter Test of Gravitational Wave Dispersion with Principal Component Analysis}

\author{Zhi-Chu Ma \inst{1,2}, Rui Niu \inst{1,2}, Wen Zhao \inst{1,2}}

\institute{CAS Key Laboratory for Research in Galaxies and Cosmology, Department of Astronomy, University of Science and Technology of China, Hefei 230026, China; \and School of Astronomy and Space Sciences, University of Science and Technology of China, Hefei, 230026, China; 
{\it nrui@mail.ustc.edu.cn}}

\date{\today}

\abstract{
In this work, we consider a conventional test of gravitational wave (GW) propagation which is based on the phenomenological parameterized dispersion relation to describe potential departures from General Relativity (GR) along the propagation of GWs. But different from tests conventionally performed previously, we vary multiple deformation coefficients simultaneously and employ the principal component analysis (PCA) method to remedy the strong degeneracy among deformation coefficients and obtain informative posteriors.
The dominant PCA components can be better measured and constrained, thus are expected to be more sensitive to potential departures from the waveform model. 
Using this method we analyze 10 selected events and get the result that the combined posteriors of the dominant PCA parameters are consistent with GR within 3-$\sigma$ uncertainty. 
The standard deviation of the first dominant PCA parameter is 3 times smaller than that of the original dispersion parameter of the leading order. 
However, the multi-parameter test with PCA is more sensitive to not only potential deviations from GR but also systematic errors of waveform models.
the difference in results obtained by using different waveform templates indicates that the demands of waveform accuracy are higher to perform the multi-parameter test with PCA.
}

\authorrunning{Z.-C. Ma, R. Niu, W. Zhao}            
\titlerunning{Multi-parameter test of GW propagation with PCA}  

\maketitle

\section{Introduction} \label{sec_introduction}
General Relativity (GR) is considered as the most successful gravity theory which has been intensively tested in the past across scales of laboratory experiments to observations of the large-scale structure of the universe \citep{Berti2015,Will2014,Hoyle2001,Adelberger2001,Jain2010,Koyama2016,Stairs2003,Manchester2015,Wex2014,Kramer2017}.
However, on the theoretical side, difficulties in problems of singularity and quantization \citep{DeWitt1967,Kiefer2007} hint at the possible incompleteness of GR. 
On the observational side, to explain current observations of galaxies and cosmology within the framework of GR, conceptions like dark matter and dark energy have to be introduced \citep{Frieman2008,Porter2011}, while there is another possibility that GR might be invalid at this scale \citep{Debono2016}. 
These facts continuously motivate people to pursue higher precision or develop new methods for performing tests on GR.
In recent years, gravitational wave (GW) observations provide an unprecedented way to test GR. 

The direct detection of the binary black hole coalescence event, GW150914 \citep{Abbott2016}, initiates the era of gravitational wave astronomy. More than 90 GW events from compact binary coalescence have been detected \citep{Abbott2019,Abbott2020,Collaboration2021e,Collaboration2021f} in the three previous observing runs of LIGO, Virgo, and KAGRA collaboration (LVK).
Thorough tests on GR have been performed by LVK based on these detected GW data \citep{Collaboration2021f,Collaboration2020,Collaboration2021g}.

The tests performed by LVK include three aspects of GW, generation, propagation, and polarization. The methods used can be classified as consistency tests and parameterized tests \citep{Collaboration2021f,Collaboration2020,Collaboration2021g}. Consistency tests aim at checking whether the observed data are consistent with predictions of GR, such as the residual test where the best-fit waveform will be subtracted from data and checking whether there is remnant coherent power in residuals \citep{Cornish2015}, or the inspiral-merger-ringdown consistency test where the low frequency and high frequency parts of signals are used to perform parameter estimation separately and checking whether the results are consistent \citep{Ghosh2017,Ghosh2016}.
While, parameterized tests adopt specific parametrization of possible deviations from GR. 
For example, the parameterized test of GW generation \citep{Li2012,Agathos2014} utilizes the parametrization based on the post-Newtonian (PN) expansion structure. Deformation coefficients are added to original PN coefficients which are solely determined by masses and spins of the binary in GR. The additional deformation coefficients will be estimated as free parameters in Bayesian parameter estimation, which can capture various possible unmodeled effects and can also be mapped to modifications in specific alternative gravity theories through the parameterized post-Einsteinian (ppE) framework \citep{Yunes2009}. 
Similarly, the parameterized dispersion relation is used in the test of GW propagation \citep{Mirshekari2012}. Additional power terms of GW momentum are added in the dispersion relation, which can make different frequency components of GWs propagate with different speeds, thus distorting the observed waveform \citep{Mirshekari2012,Will1998}. 

There are multiple deformation coefficients in parameterized tests for capturing various potential deviations.
Due to the limitation of sensitivity of current detectors, if varying all deformation coefficients simultaneously in parameter estimation, correlations among parameters will lead to uninformative posteriors. 
Thus, in tests performed by LVK, only one deformation parameter is allowed to vary at a time \citep{Abbott2016a,Abbott2019c,Collaboration2021f,Collaboration2020,Collaboration2021g}.

However, in the most general and agnostic-priori case, all deformation parameters need to be considered as free parameters in Bayesian inference. The class of parameterized tests where multiple deformation parameters are constrained simultaneously is referred as multi-parameter tests \citep{Gupta2020,Datta2020,Saleem2021}. 
Previous works \citep{Gupta2020,Datta2020} show that the multi-band observations by third-generation ground-based detectors and space-borne detectors can allow people to perform multi-parameter tests and get tight constraints on PN deformation coefficients.
Another approach to perform multi-parameter tests is employing the method of Principal Component Analysis (PCA) to reduce correlations among parameters and get informative posteriors \citep{Ohme2013,Pai2012,Saleem2021,Datta2022,Datta2023,Shoom2023}.
The previous work \citep{Saleem2021} applies this method in the parameterized test of GW generation. In this work, we will extend this method to the parameterized test of GW propagation.
The rest of this paper is organized as follows.
In Section \ref{sec_method}, we introduce the paremeterized test of GW propagation, and the method of PCA for multi-parameter tests. We apply this method to 10 selected GW events, and present results in Section \ref{sec_results}. The summary is shown in Section \ref{sec_summary}.

\section{Methods} \label{sec_method}

\subsection{The multi-parameter test of GW propagation}
The main effort of this paper is extending the multi-parameter test of GW generation with PCA performed in the previous work \citep{Saleem2021} to the test of propagation. 
We focus on the effects of dispersion, while other propagation effects like birefringence \citep{Wang2021g,Wang2022b,Niu2022,Zhao2020a,Okounkova2021} or amplitude damping \citep{Nishizawa2018,Belgacem2018} are not included in this paper.
The test of propagation considered here is based on the phenomenological modified dispersion relation \citep{Mirshekari2012} which reads 
\begin{equation}
E^2=p^2 + \sum_\alpha A_\alpha p^\alpha,
\end{equation}
where $E$ and $p$ are the energy and momentum of GW respectively, $A_\alpha$ is the free parameters for capturing various potential deviations from GR, and $\alpha$ corresponds to modifications at different frequency orders. 
When $A_\alpha=0$ for all $\alpha$, this modified dispersion relation returns to the case of GR.
Leading-order modifications in various alternative theories can be mapped to the terms with different values of $\alpha$ \citep{Yunes2016,Mirshekari2012,Yunes2009}.

This phenomenological modified dispersion relation is a generic framework and can cover a wide variety of alternative theories, where the behaviors of GW propagation are different with GR, including theories with massive gravitons or various Lorentz-violating alternative gravity theories \citep{Mirshekari2012}. For example, the case of $\alpha=0$ and $A_0 > 0$ corresponds to deformations induced by massive gravitons \citep{Will1998}; the leading-order corrections in multi-fractal spacetime \citep{Calcagni2010} and the doubly special relativity \citep{AmelinoCamelia2002} can be mapped to the term of $\alpha = 2.5$ and 3; The Hor\v{a}va-Lifshitz theory \citep{Horava2009}, extra-dimensional theories \citep{Sefiedgar2011}, and the standard model extension where if only non-birefringent effects are considered \citep{Kostelecky2016}, have leading modifications with $\alpha=4$.

Additional power terms of momentum in the modified dispersion relation will make different frequency components of GW propagate with different speeds, thus leaving imprints in the waveform observed. 
Assuming the waveform in the local wave zone is consistent with GR, the additional corrections in phase induced by dispersion effects along propagation are given by \citep{Mirshekari2012}
\begin{equation} \label{eq_modefication_original}
\delta \Phi_\alpha(f)=\text{sign}(A_\alpha)\left\{ 
\begin{aligned} 
&\frac{\pi (1+z)^{\alpha-1}D_\alpha}{\alpha-1}\lambda_\alpha^{\alpha-2}f^{\alpha-1}, & \alpha \neq 1, \\
&\frac{\pi D_\alpha}{\lambda_\alpha}\ln\left(\pi \mathcal{M}f\right), &\alpha=1.\end{aligned} \right. 
\end{equation}
In the above equation, $\mathcal{M}$ is the chirp mass of the binary, and $\lambda_\alpha=h|A_\alpha|^{1/(\alpha-2)}$ with the Planck constant $h$, $D_\alpha$ is defined as 
\begin{equation}
D_\alpha=\frac{(1+z)^{1-\alpha}}{H_0}\int_0^z \frac{(1+z')^{\alpha-2}}{\sqrt{\Omega_{\text{m}}(1+z')^3+\Omega_\Lambda}}dz', 
\end{equation}
where $H_0$ is the Hubble constant, $z$ is the redshift of the source, $\Omega_{\text{m}}$ and $\Omega_\Lambda$ are the matter and dark energy density parameters.
We use the values reported in Planck 2018 results \citep{Aghanim2020}, where $H_0=67.66 \ \text{km} \ \text{s}^{-1} \ \text{Mpc}^{-1}$, $\Omega_\text{m}=0.3111$ and $\Omega_\Lambda=0.6889$.

For the convenience of varying all deformation coefficients simultaneously in parameter estimation, we consider a parameterization which is slightly different with that used by LVK \citep{Abbott2016a,Abbott2019c,Collaboration2021f,Collaboration2020,Collaboration2021g}. 
Dimensionless coefficients 
\begin{equation} 
    \delta\phi_\alpha = 1\ {\rm Gpc} (10^3 \ M_\odot)^{1-\alpha} \lambda_\alpha^{\alpha-2},
\end{equation}
are constructed, where the factors of distance and mass in front are adjusted according to typical values of detected sources in order to scale the magnitude of deformation coefficients to a close order.
Therefore, the phase correction of Eq. \ref{eq_modefication_original} can be rewritten by
\begin{equation} 
    \delta\Phi_\alpha(f)=\text{sign}(A_\alpha)
    \left\{\begin{aligned}
    &\delta\phi_\alpha
    \frac{(1+z)^{\alpha-1}}{(\alpha-1)\pi^{\alpha-2}}
    \left(\frac{10^3 \ M_\odot}{M}\right)^{\alpha-1}
    \left(\frac{D_\alpha}{1 \ \mathrm{Gpc}}\right)
    \left(\pi \mathcal{M} f\right)^{\alpha-1},&\alpha\neq1,
    \\
    &\delta\phi_\alpha\frac{D_\alpha}{1 \ \mathrm{Gpc}}\pi \ln(\pi \mathcal{M}f) , &\alpha=1.
    \end{aligned} \right.
\end{equation}

The above modifications are added into a GR waveform model as the template used in the Bayesian parameter estimation. 
The measurement of all free parameters in the template, including deformation coefficients of dispersion effects and source properties in GR, will be presented through posterior distributions obtained by the nested sampling \citep{Skilling2004,Skilling2006} or the Markov chain Monte Carlo \citep{Hastings1970,ForemanMackey2013} algorithms.
The posterior samples of $\delta \phi_\alpha$, after marginalizing over GR parameters, are manipulated by the PCA method to reduce correlations among these coefficients and obtain informative posteriors.
We will describe above procedures in more detail in the following.

\subsection{Bayesian parameter estimation with GWs}
Next, we present a brief review of Bayesian parameter estimation for GW data \citep{Abbott2020c}.
The physical information is extracted from observations through the Bayesian framework which estimates the probability distributions of parameters $\boldsymbol{\theta}$ in the model $M$ according to observed data $\boldsymbol{d}$. 
The measurement is encoded by posterior distributions $p(\boldsymbol{\theta}|\boldsymbol{d}, M)$.
According to Bayes' theorem, the posterior is given by 
\begin{equation}
p(\boldsymbol{\theta}|\boldsymbol{d}, M) = \frac{p(\boldsymbol{\theta})p(\boldsymbol{d}|\boldsymbol{\theta}, M)}{p(\boldsymbol{d})}.
\end{equation}
Here, $p(\boldsymbol{\theta})$ is the prior probability which reflects our prior knowledge of the parameters ahead of observations, 
$p(\boldsymbol{d}|\boldsymbol{\theta}, M)$ is the likelihood whose value is the probability of the occurrence of a noise realization that is just equal to the observed data $\boldsymbol{d}$ subtracting a GW signal given by the model $M$ with the specific values of $\boldsymbol{\theta}$,
$p(\boldsymbol{d})$ is the normalizing factor and also referred as evidence for its usage in model selection.

In the GW data analyses context, assuming the noise is Gaussian and stationary, the likelihood can be written as 
\begin{equation}
p(\boldsymbol{d}|\boldsymbol{\theta}, M) \propto 
\exp 
\left[-\frac{(\boldsymbol{d} - \boldsymbol{h}|\boldsymbol{d} - \boldsymbol{h})}{2} \right].
\end{equation}
Here, $\boldsymbol{h}$ is the GW signal given by a set of parameters $\boldsymbol{\theta}$ in a specific model $M$, the brackets denote the inner product which is defined as
\begin{equation}
(a|b) = 4\text{Re}\int df\frac{a^*(f)b(f)}{S_n(f)}, 
\end{equation}
where $S_n(f)$ is the power spectral density (PSD) of detector noise.
For GW signals, the dimension of parameter space is usually high, and the likelihood evaluation is usually computational expensive.
Computing posteriors on a grid is impractical.
Stochastic sampling algorithms, including nested sampling \citep{Skilling2004,Skilling2006} or the Markov chain Monte Carlo \citep{Hastings1970,ForemanMackey2013}, are used to obtain random samples whose densities can be approximated to the posterior probabilities.

In our situation, the model contains multiple dispersion parameters $\delta \phi_\alpha$ together with GR parameters. Different with tests performed by LVK, we vary multiple deformation parameters simultaneously in the parameter estimation introduced above.
After marginalizing GR parameters, the posterior samples of dispersion parameters are manipulated by the procedure discussed in the next subsection to reduce correlations.

\subsection{Principle component analysis} \label{subsec_pca}
Different with the tests performed by LVK, multiple dispersion parameters are varied simultaneously in our parameter estimation. Due to correlations among these deformation parameters, varying multiple parameters will lead to less informative posteriors.
However, previous works \citep{Ohme2013,Pai2012,Saleem2021,Datta2022,Datta2023,Shoom2023,Niu2024} show that this problem can be remedied by the method of PCA.

The method of PCA will find a new set of bases for the parameter space by a linear combination of original parameters.
The new constructed parameters have reduced correlations, and can be better constrained by data.
To obtain the new bases, we first need to compute the covariance matrix $\boldsymbol{\Sigma}$ for the posterior samples of dispersion parameters $\delta\phi_\alpha$ after marginalizing GR parameters, which is given by 
\begin{equation}
\Sigma_{ij} = \Bigl\langle 
\bigl( \delta \phi_i-\langle \delta \phi_i\rangle \bigr)  
\bigl( \delta \phi_j-\langle \delta \phi_j\rangle \bigr) 
\Bigr\rangle.
\end{equation}
Here, $\langle\cdots\rangle$ denotes the expectation value.
Then, diagonalizing the covariance matrix $\boldsymbol{\Sigma}$ and it can be written in terms of eigenvectors and eigenvalues as
\begin{equation}
\boldsymbol{\Sigma}=\boldsymbol{U} \boldsymbol{\Lambda}\boldsymbol{U}^T. 
\end{equation}
$\boldsymbol{\Lambda}$ is a diagonal matrix whose diagonal elements are eigenvalues of $\boldsymbol{\Sigma}$. $\boldsymbol{U}$ is a matrix composed by eigenvectors of $\boldsymbol{\Sigma}$.
The eigenvectors are the new bases we are looking for, in which each dimension is linearly uncorrelated.
The posteriors on this new set of bases can be obtained  by the transformation
\begin{equation} 
\delta \phi^{\text{PCA}}_{i\text{-th}}=\sum_j U^{ij}\delta \phi_j,
\end{equation}
where $\delta \phi^{\text{PCA}}_{i\text{-th}}$ denoted the $i$-th principal component. 

The levels of how principal the component is are indicated by the values of corresponding eigenvalues.
The eigenvector with the smallest eigenvalue is the most dominant component corresponding to the new PCA parameter with the smallest error bar.
The method of PCA makes the information of posteriors redistribute among new bases, and collects information of all PN orders into dominant components.
Therefore, potential deviations can be more obvious when showing posteriors by PCA parameters.

\subsection{Combination of results from multiple events} \label{subsec_combine}
According to the method introduced above, we can perform the multi-parameter test of GW dispersion using a single GW event. In this subsection, we introduce the operation for combining information from multiple events.

The deviations from GR if exist are believed to be universal for all events.
Assuming the observation of each event is an independent measurement of deformation parameters $\delta\phi_\alpha$, we can directly multiply the posterior probabilities of $\delta\phi_\alpha$ given by each event to get the combined posteriors, which is the same as in parameterized tests performed by LVK \citep{Abbott2016a,Abbott2019c,Collaboration2021f,Collaboration2020,Collaboration2021g}.
However, since the linear combination of original dispersion parameters to get PCA parameters $\delta \phi^{\text{PCA}}_{i\text{-th}}$ is unique for each event, it is inappropriate to multiply posteriors of PCA parameters.
Therefore, we resample the combined posteriors of $\delta\phi_\alpha$ and perform a separate operation of PCA to obtain the combined posteriors of $\delta \phi^{\text{PCA}}_{i\text{-th}}$.

In practice, we use Gaussian kernel density estimation (KDE) to obtain posterior probabilities of $\delta\phi_\alpha$ from its posterior samples for each event.
Then, using a new MCMC sampling to sample the cumulative production of KDEs of all events, we can obtain the posterior samples of $\delta\phi_\alpha$ combining information from all events.
The combined posteriors of PCA parameters $\delta \phi^{\text{PCA}}_{i\text{-th}}$ are obtained by performing a separate operation of PCA on the combined posterior samples of $\delta\phi_\alpha$

As will be discussed in the next section, to verify the robustness of using different waveform models, we do the same parameter estimation with different templates for each event.
We use the method discussed in \citep{Ashton2019a} to average the results obtained by using different waveform approximations, where the posterior for a single event estimated by using a set of models $M_i$ is given by
\begin{equation}
    p(\boldsymbol{\theta}|\boldsymbol{d}, \{M_i\}) = \sum_i p(\boldsymbol{\theta}|\boldsymbol{d}, M_i) \xi_i,
\end{equation}
where the weights $\xi_i$ are defined as $\xi_i = Z_i/\sum_j Z_j$ with $Z_i$ denoting the evidence of each waveform model.

\section{Results and discussions} \label{sec_results}

We apply the method elaborated in the last section on current observed GW data and present results in this section.
Due to the limitation of sensitivity of current detectors, Following the multi-parameter test of generation performed in \citep{Saleem2021}, we also do not consider the full set of dispersion parameters.
We only include the case of $\alpha =$ 0, 2.5, 3, and 4 in our analyses considering the examples of leading order modifications in alternative theories mentioned in \citep{Abbott2019b,Mirshekari2012}.

We analyse the 10 farthest events detected by LVK in three previous runs with criteria of that the false alarm rate (FAR) less than $10^{-3}\ \text{year}^{-1}$, the signal-to-noise ratio (SNR) greater than 12, and the probability of astrophysical origin ($p_{\text{astro}}$) greater than 0.99.
The information of selected events is summarized in Table \ref{tab_events}.
We obtain strain data and noise PSD data from public data release of LVK\footnote{\url{https://gwosc.org/eventapi/html/GWTC/}}.
The strain data with glitch subtraction are used in parameter estimation for events GW190503\_185404 and GW190513\_205428.

We use \texttt{bilby} \citep {Ashton2019} with nested sampler \texttt{pymultinest} \citep{Buchner2014,Feroz2008,Feroz2007} to perform Bayesian parameter estimation.
Since the method of PCA is more sensitive to not only possible deviations from GR but also systematic errors of waveform template, we also use different waveform including \texttt{SEOBNRv4\_ROM} \citep{Puerrer2015,Puerrer2014}, \texttt{IMRPhenomXAS} \citep{Pratten2020}, \texttt{IMRPhenomXHM} \citep{Pratten2021}, and \texttt{NRSur7dq4} \citep{Varma2019}, in our parameter estimation.

\begin{figure}
    \centering
    \includegraphics[width=\columnwidth]{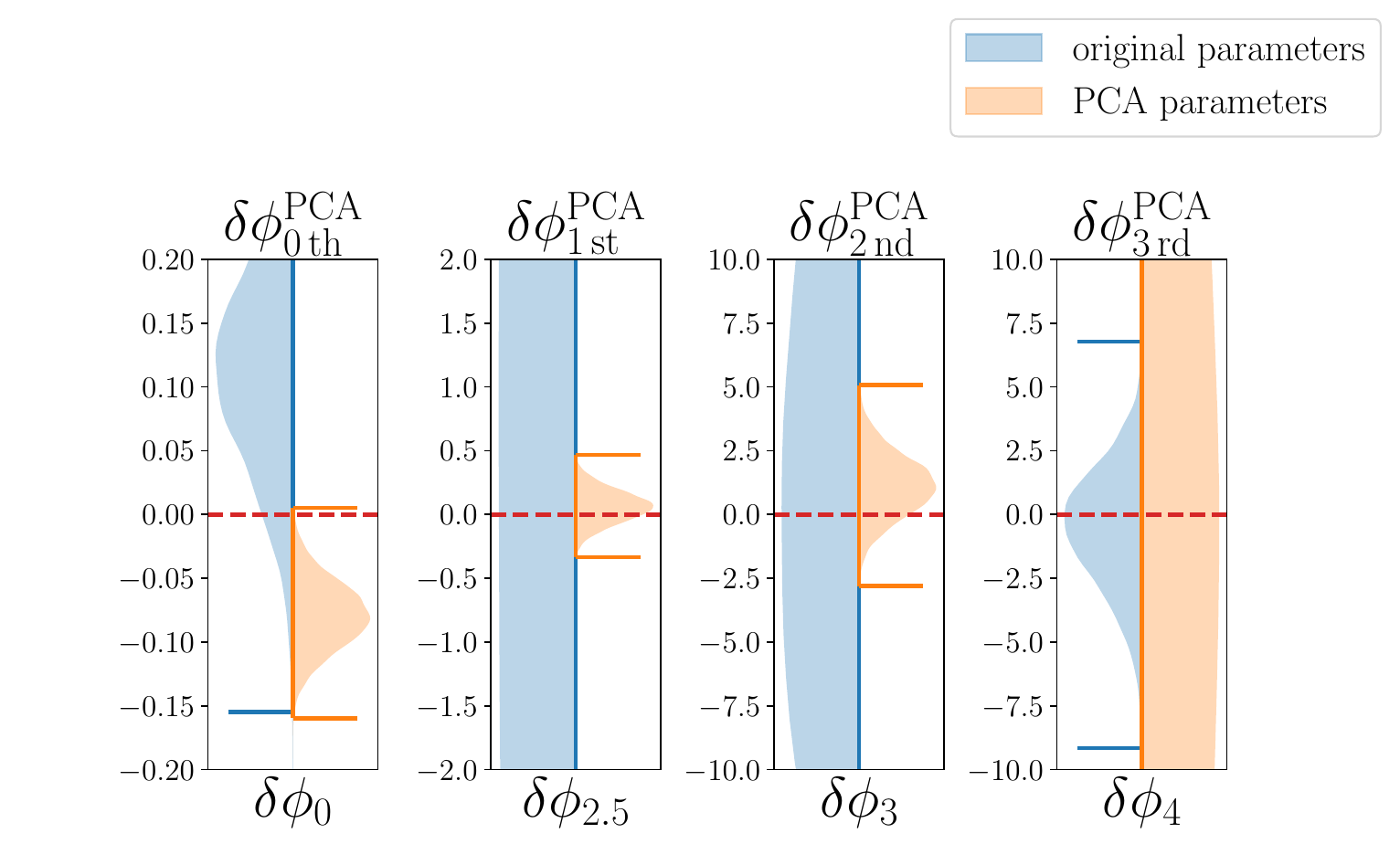}
    \caption{\textbf{Posteriors of deformation parameters before and after performing PCA.} The result shown here is obtained by combining posteriors of all selected events and averaging with weights of evidence for different waveform templates. The posteriors of original dispersion parameters are shown at left side with blue shadows, and posteriors of PCA parameters are shown at right side with orange shadows. The solid lines denote 3-$\sigma$ error bars and the red dashed lines denote GR values.}
    \label{fig_violin_combining}
\end{figure}

\begin{figure}
    \centering
    \includegraphics[width=0.49\columnwidth]{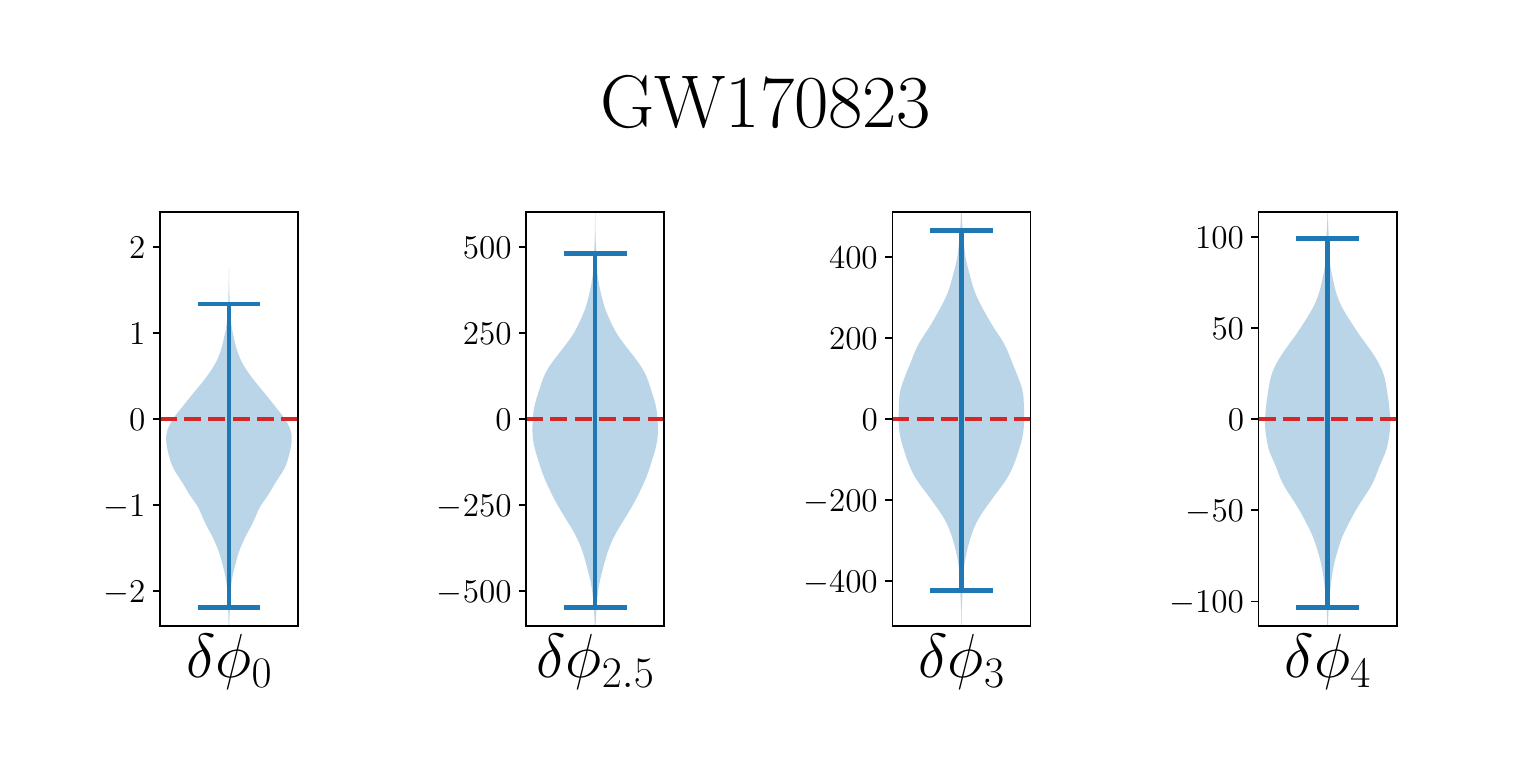}
    \includegraphics[width=0.49\columnwidth]{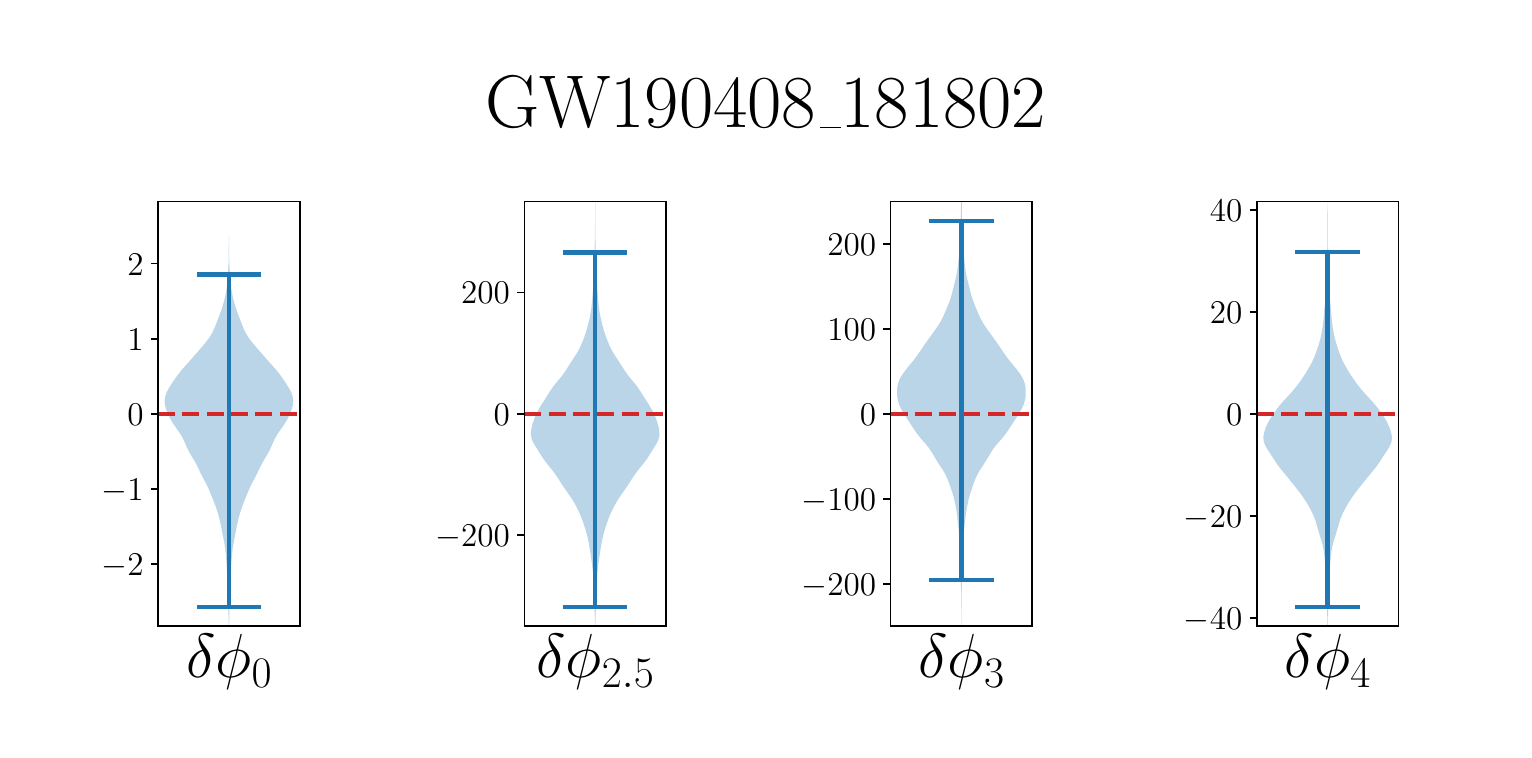}
    \includegraphics[width=0.49\columnwidth]{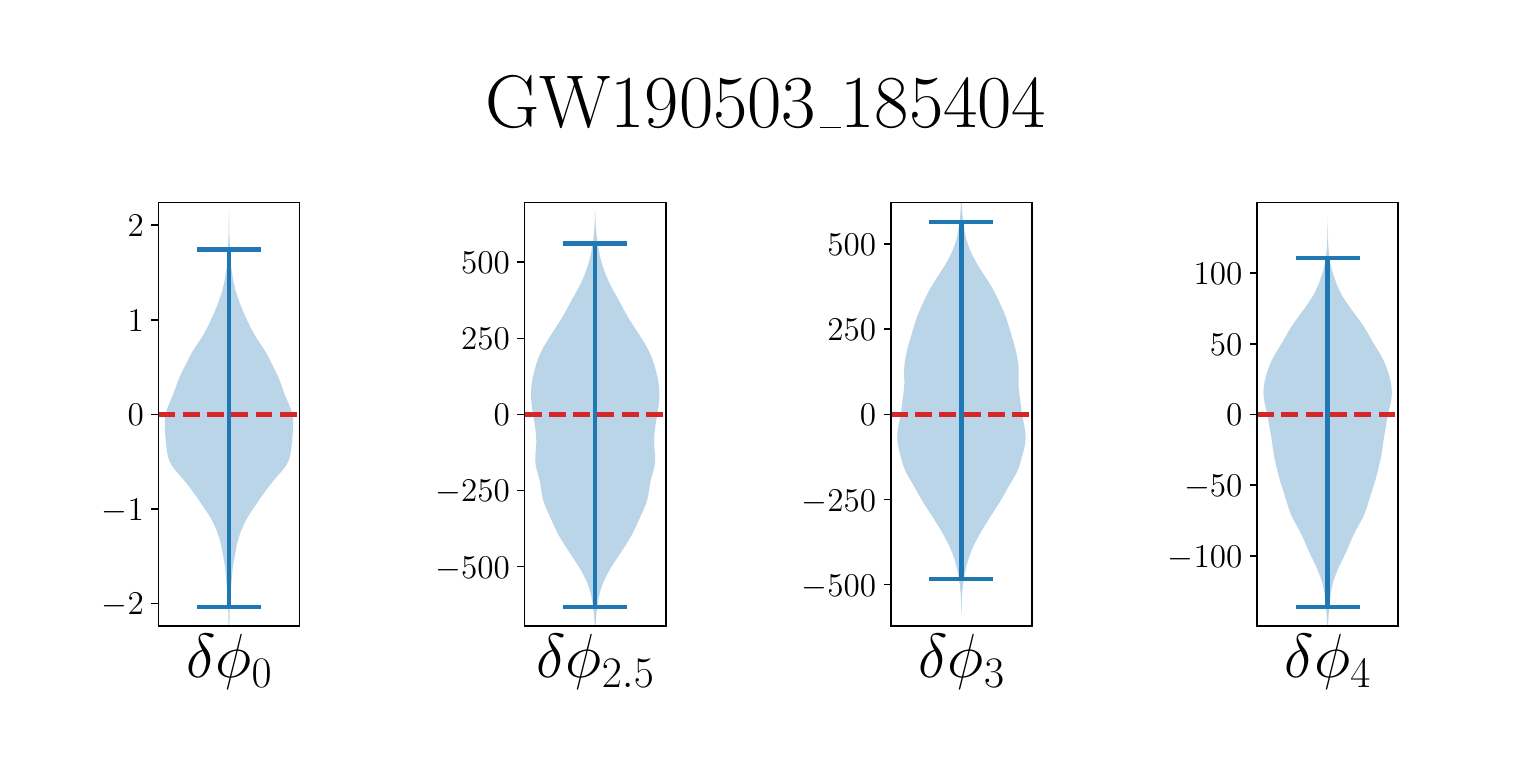}
    \includegraphics[width=0.49\columnwidth]{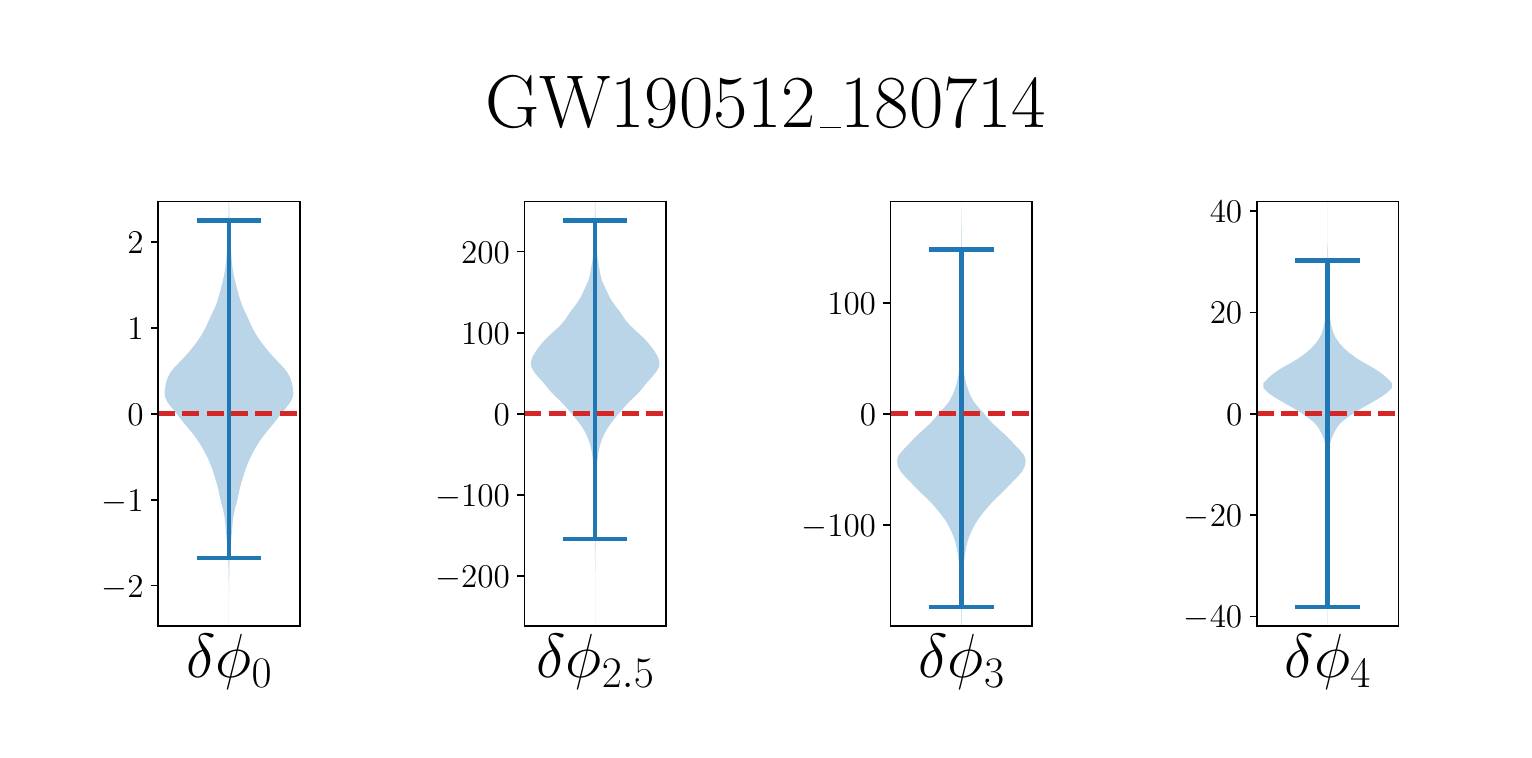}
    \includegraphics[width=0.49\columnwidth]{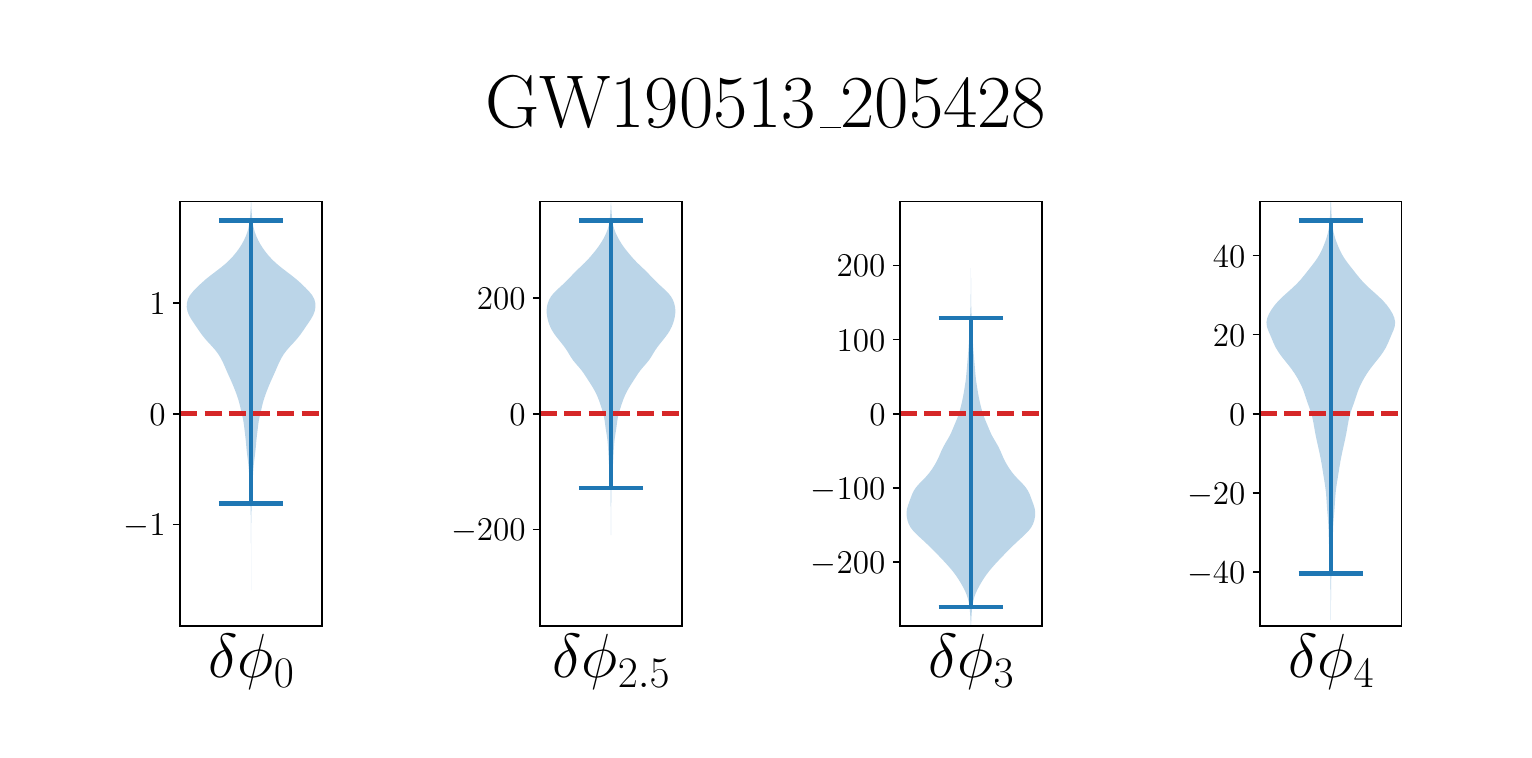}
    \includegraphics[width=0.49\columnwidth]{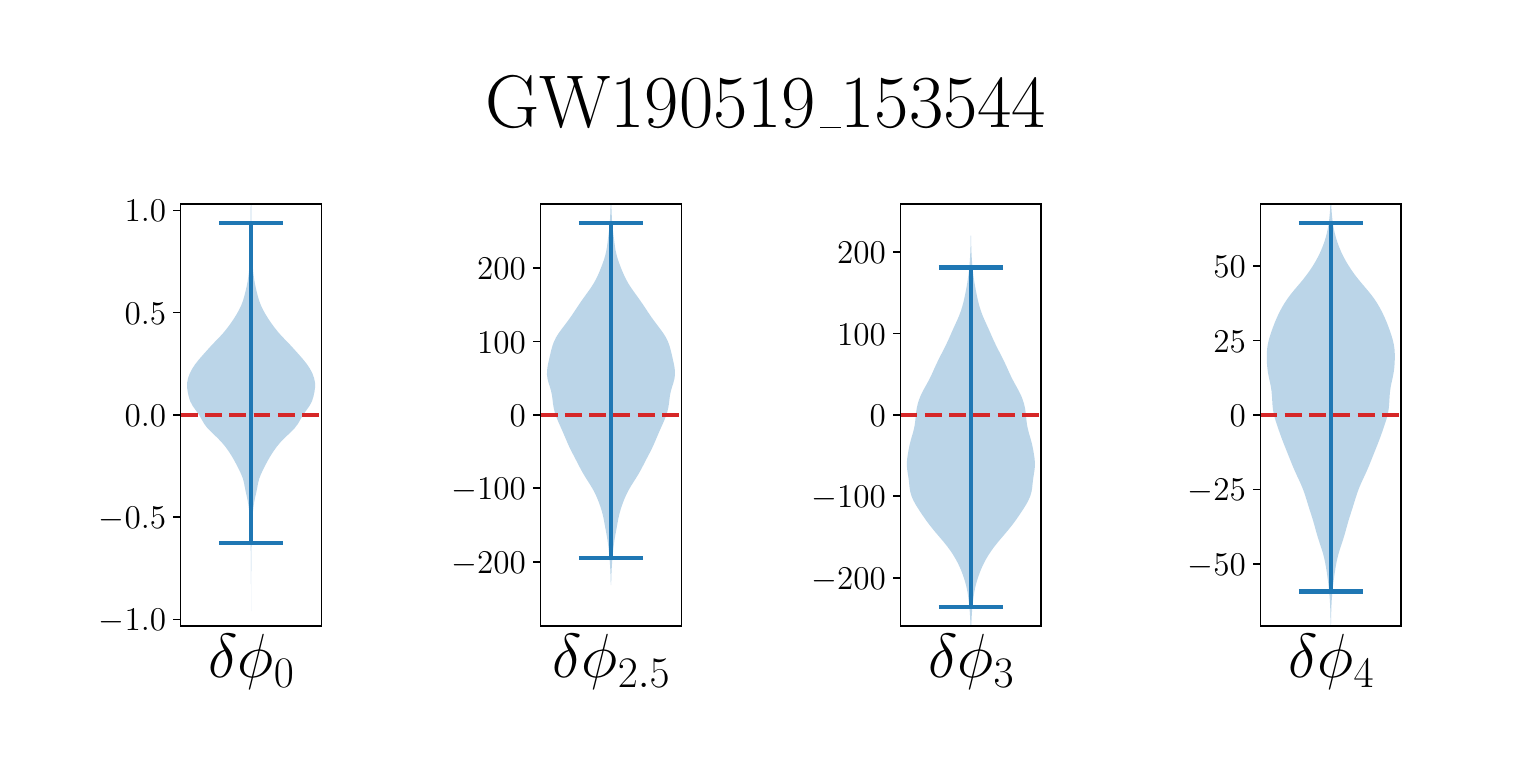}
    \includegraphics[width=0.49\columnwidth]{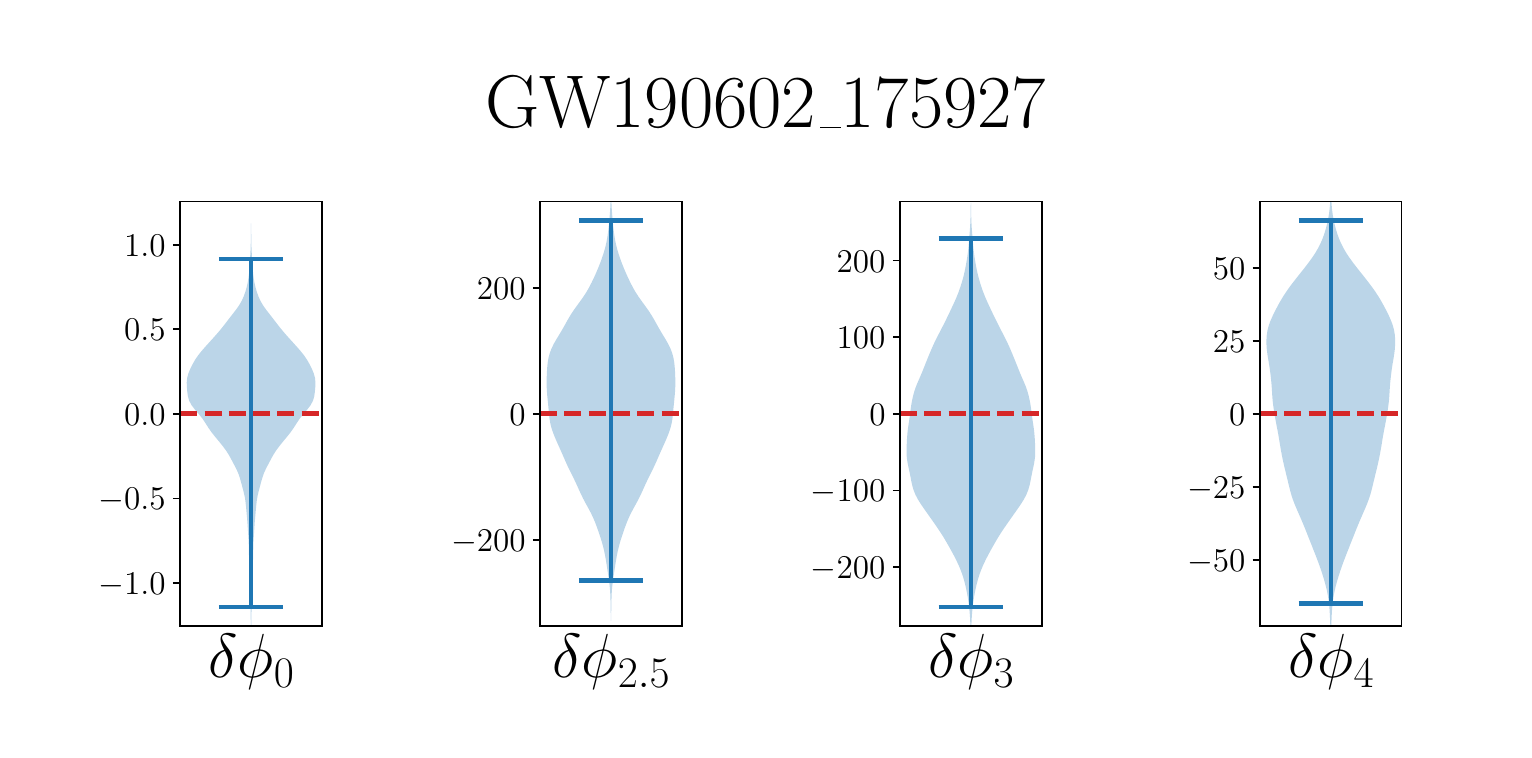}
    \includegraphics[width=0.49\columnwidth]{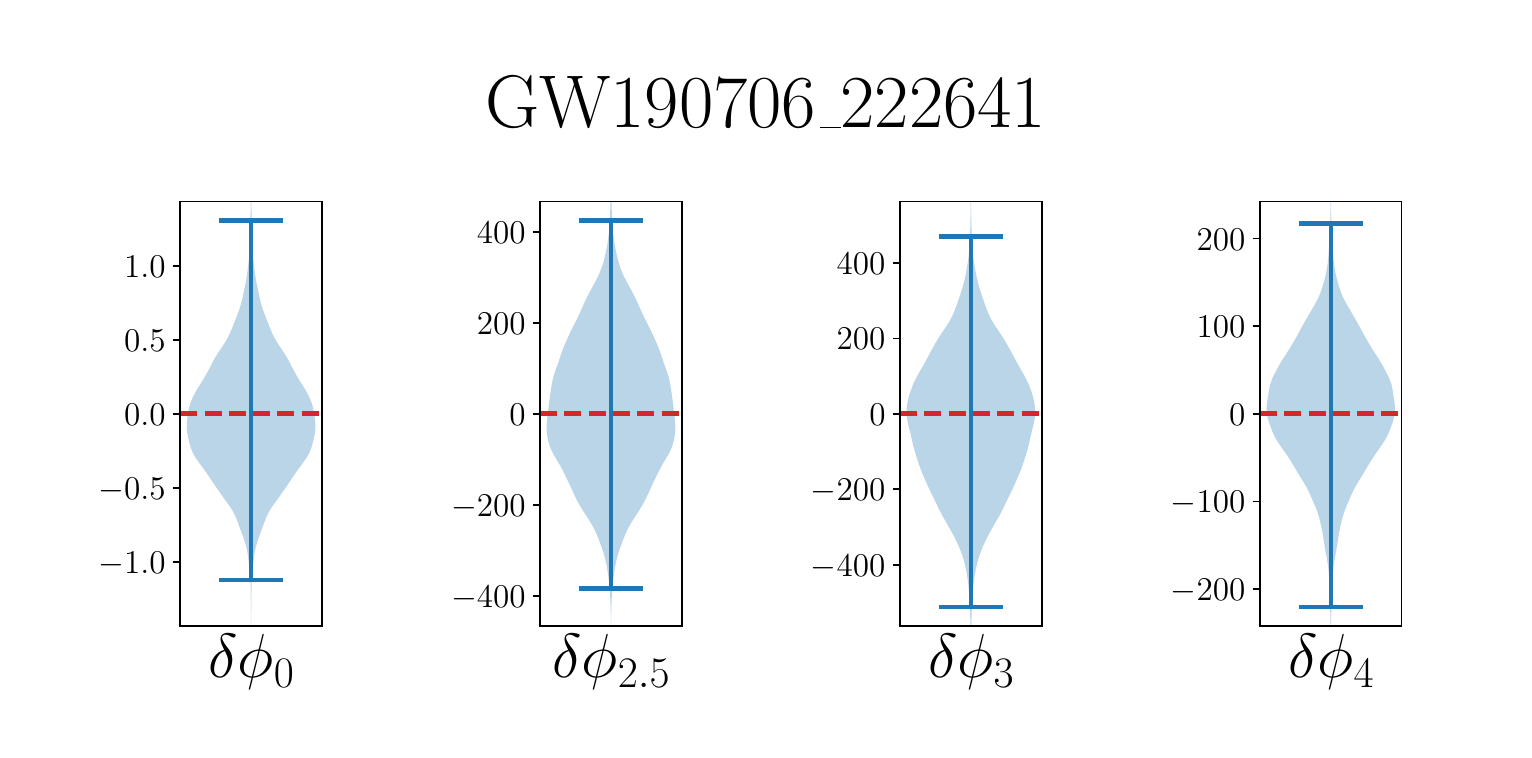}
    \includegraphics[width=0.49\columnwidth]{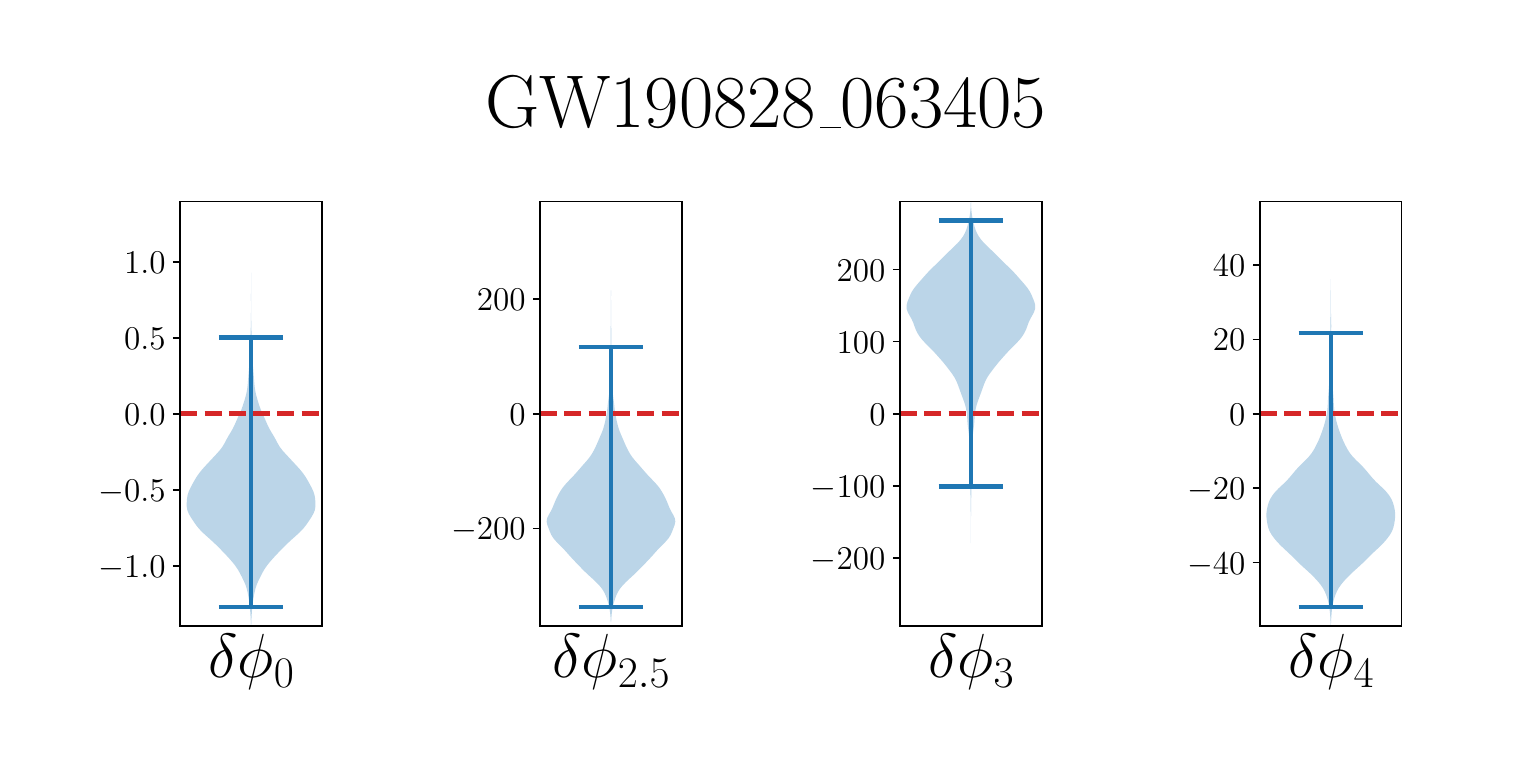}
    \includegraphics[width=0.49\columnwidth]{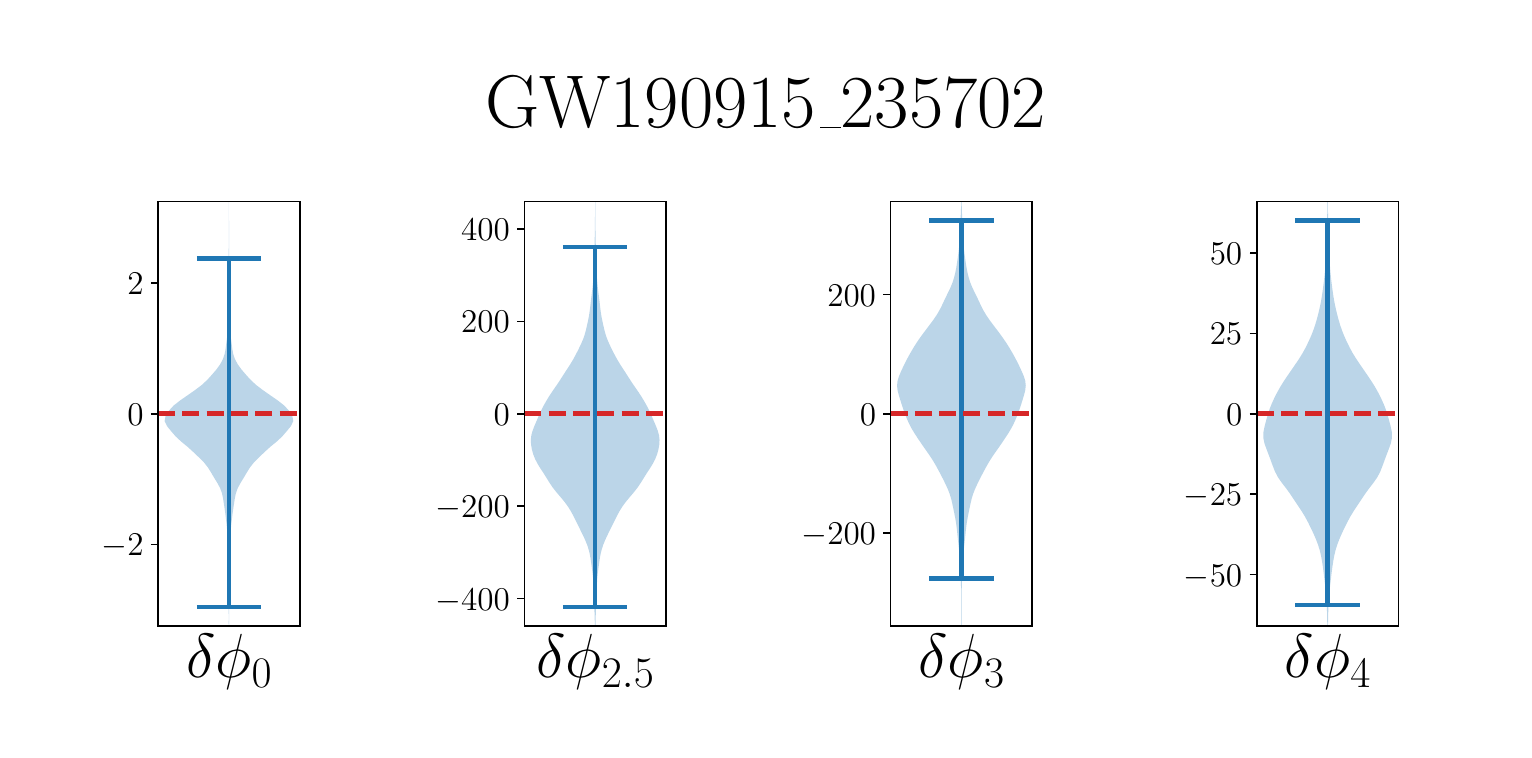}
    \caption{\textbf{Posterior distributions of original dispersion parameters for each selected event.}
    GR values are denoted by red dashed lines, and 3-$\sigma$ uncertain ranges are denoted by blue solid lines. the results shown here have been averaged with the weights of evidence of different waveform models.}
    \label{fig_each_events}
\end{figure}

The final results are presented in Figure \ref{fig_violin_combining} where the posteriors are obtained by averaging results of different waveform models with weights of evidence and combining results of all selected events using the method discussed in Section \ref{subsec_combine}, and the results of each event are also shown in Figure \ref{fig_each_events}.
The posteriors of original dispersion parameters $\delta\phi_\alpha$ are shown at left sides with blue shadow in each subplot, and posteriors of PCA parameters $\delta \phi^{\text{PCA}}_{i\text{-th}}$ are shown at right sides with orange shadow.
The blue and orange solid lines denote the 3-$\sigma$ error bars, and the red dashed lines denote the GR values.
Due to the limitation of plot ranges, the error bars of $\delta\phi^{\text{PCA}}_{3\text{rd}}$, $\delta\phi_{2.5}$, $\delta\phi_{3}$, and the upper limit of $\delta\phi_0$ are out of the visible ranges.
The explicit values of error bars and departures of maximum likelihood values from GR values are summarized in Table \ref{tab_results} for quantitative comparison.

We can observe that the posteriors of all parameters, including original dispersion parameters and PCA parameters, are consistent with GR within 3-$\sigma$ uncertain ranges through Figure \ref{fig_violin_combining} and Table \ref{tab_results}.
As discussed in Section \ref{subsec_pca}, The dominant PCA parameters can be better constrained thus have smaller error bars, which are obviously shown in Figure \ref{fig_violin_combining} observing the first three components. For example, the range of 3-$\sigma$ uncertainty of the first dominant PCA parameter is 3 times smaller than that of the original dispersion parameter of the leading order.
We can also notice that the maximum likelihood values are not exactly the GR values, which is reasonable considering unavoidable errors in waveform models and non-perfect Gaussian noise in reality.
These departures are more obvious in dominant PCA parameters.
Observing the subplot of the first dominant component $\delta\phi^{\text{PCA}}_{0\text{-th}}$ and the leading order dispersion parameters $\delta\phi_{0}$, the departure between the maximum likelihood value and the GR value of $\delta\phi^{\text{PCA}}_{0\text{-th}}$ is 2.85 $\sigma$, while this value is 1.34  $\sigma$ for $\delta\phi_{0}$.
This in one hand shows that the dominant PCA parameter is more sensitive to any departures in posteriors, and on the other hand indicates the demand of waveform accuracy is higher to search deviations using PCA.

\begin{figure}
    \centering
    \includegraphics[width=\columnwidth]{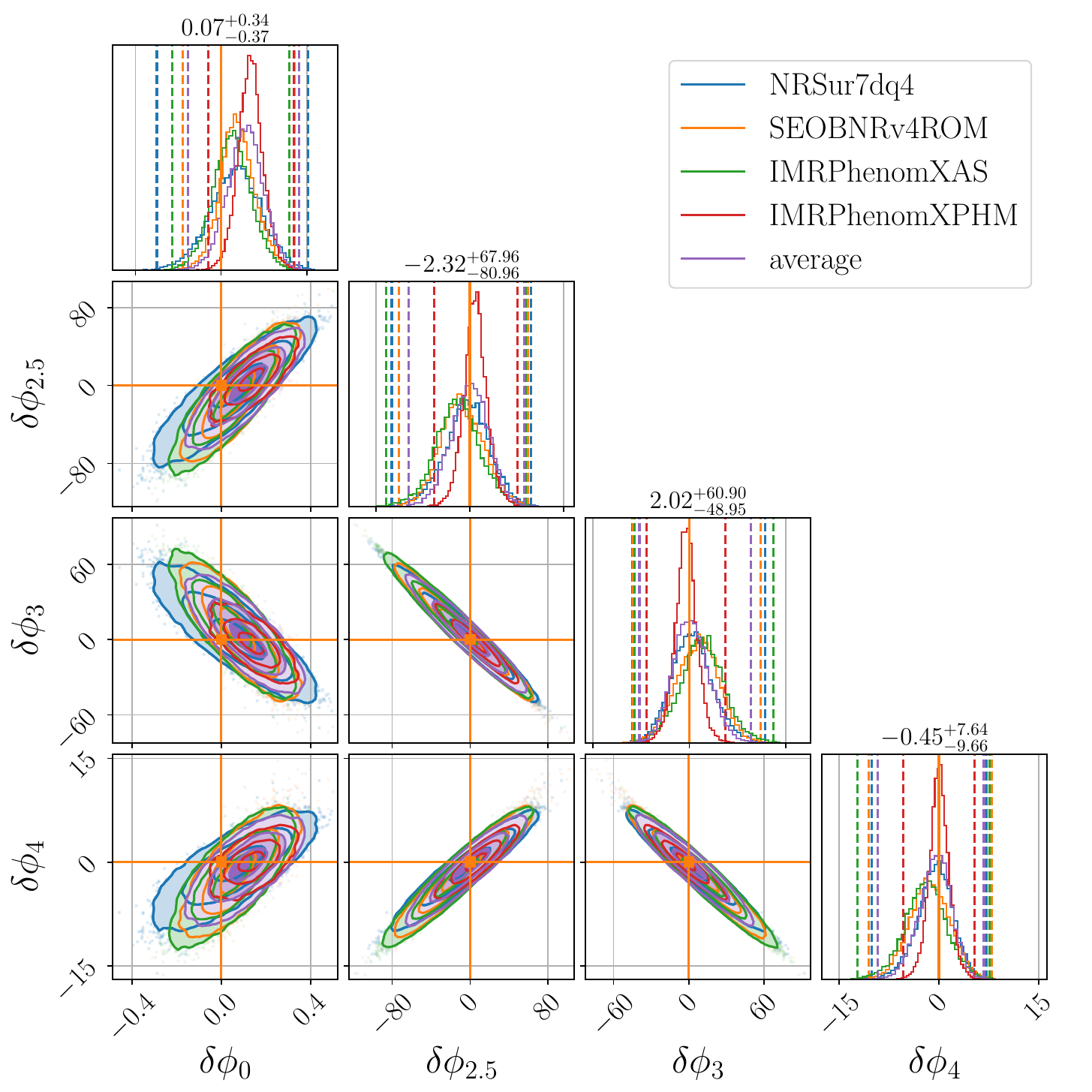}
    \caption{\textbf{Posteriors of original dispersion parameters, which are combined results of all selected events, obtained by using different waveform models. }
    The orange solid lines denote GR values and the dashed lines denote 3-$\sigma$ uncertain ranges. 
    }
    \label{fig_corner}
\end{figure}

\begin{figure}
    \centering
    \includegraphics[width=\columnwidth]{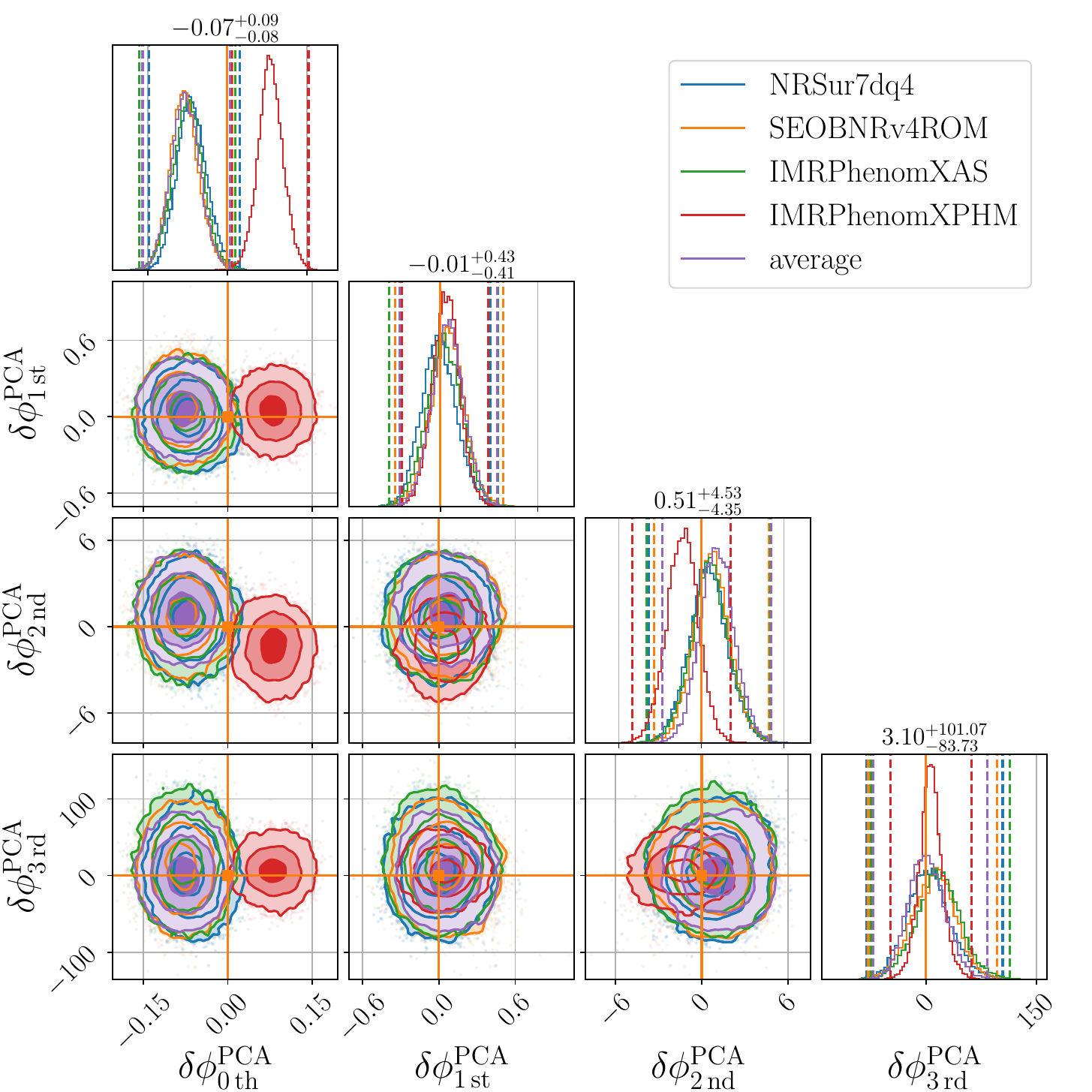}
    \caption{\textbf{The results of applying PCA on the posteriors shown in Figure \ref{fig_corner}}.
    Same as Figure \ref{fig_corner}, the orange solid lines denote GR values and dashed lines denote 3-$\sigma$ uncertain ranges.  
    Comparing with Figure \ref{fig_corner}, we can find that correlations among these parameters are reduced, and the first 3 dominant components are better constrained.
    We can also observe that the slight difference in posteriors of original dispersion parameters can lead to significant differences in results after applying PCA, especially in the first component, which hints that the multi-parameter test with PCA may require higher accuracy of waveform templates. }
    \label{fig_corner_pca}
\end{figure}

The method of PCA is more sensitive to any departures from the considered waveform model, including both potential deviations from GR and systematic errors of the template or unmodeled effects.
Therefore, we use different waveform models in parameter estimation and the obtained results are shown in Figure \ref{fig_corner} and \ref{fig_corner_pca} for posteriors of original dispersion parameters and PCA parameters.
From these two corner plots, we can find that as introduced in Section \ref{subsec_pca}, the new bases constructed by PCA have less correlations.
Additionally, we can also observe that the result given by \texttt{IMRPhenomXPHM}, which is a phenomenological waveform model incorporating effects of higher modes and precession induced by in-plane spins, have slight difference with results given by other waveform templates in posteriors of original dispersion parameters.
But after the operation of PCA, the result of \texttt{IMRPhenomXPHM} has obvious differences with others especially in the first dominant component.
This in another aspect shows that the operation of PCA is sensitive to minor changes in posteriors of original parameters, which increases the demand for waveform accuracy to use the method of PCA.

\begin{table}[]
    \centering
    \begin{tabular}{rllll}
    \toprule
     {\bf Name}  & {\bf $D_{\text{L}}$ (Mpc)} &    SNR &  FAR ($\text{y}^{-1}$) & $p_{\text{astro}}$  \\
    \midrule
       GW170823  & $1940^{+970}_{-900}$       &   $12.2^{+0.2}_{-0.3}$ &    $\le 10^{-7}$      &  1.00  \\
GW190408\_181802 & $1540^{+440}_{-620}$       &   $14.6^{+0.2}_{-0.3}$ &    $\le 10^{-5}$      &  1.00  \\
GW190503\_185404 & $1520^{+630}_{-600}$       &   $12.2^{+0.2}_{-0.4}$ &    $\le 10^{-5}$      &  1.00  \\
GW190512\_180714 & $1460^{+510}_{-590}$       &   $12.7^{+0.3}_{-0.4}$ &    $\le 10^{-5}$      &  1.00  \\
GW190513\_205428 & $2210^{+990}_{-810}$       &   $12.5^{+0.3}_{-0.4}$ &    $1.3\times10^{-5}$ &  1.00  \\
GW190519\_153544 & $2600^{+1720}_{-960}$      &   $15.9^{+0.2}_{-0.3}$ &    $\le 10^{-5}$      &  1.00  \\
GW190602\_175927 & $2840^{+1930}_{-1280}$     &   $13.2^{+0.2}_{-0.3}$ &    $\le 10^{-5}$      &  1.00  \\
GW190706\_222641 & $3630^{+2600}_{-2000}$     &   $13.4^{+0.2}_{-0.4}$ &    $5.0\times10^{-5}$ &  1.00  \\
GW190828\_063405 & $2070^{+650}_{-920}$       &   $16.5^{+0.2}_{-0.3}$ &    $\le 10^{-5}$      &  1.00  \\
GW190915\_235702 & $1750^{+710}_{-650}$       &   $13.1^{+0.2}_{-0.3}$ &    $\le 10^{-5}$      &  1.00  \\
    \bottomrule
    \end{tabular}
    \caption{\textbf{The information of 10 selected events considered in our analyses.} 
    These events are 10 farthest events with criteria of FAR $< 10^{-3}\ \text{y}^{-1}$, SNR $>12$, and $p_{\text{astro}}>0.99$ in the Gravitational Wave Transient Catalog (GWTC).
    The values come from \href{https://gwosc.org/eventapi/html/GWTC/}{gwosc.org} which may be different with the values shown in publications of LVK \citep{Abbott2019,Abbott2020,Collaboration2021e,Collaboration2021f}.}
    \label{tab_events}
\end{table}

\begin{table}[]
    \centering
    \begin{tabular}{llllll}
        \toprule
        \multicolumn{3}{c}{\bf original dispersion parameters} & \multicolumn{3}{c}{\bf PCA parameters} \\
        \cmidrule(lr){1-3}  \cmidrule(lr){4-6}
                          & \makecell{3-$\sigma$ ranges} & \makecell{GR value at}  & &\makecell{3-$\sigma$ ranges} & \makecell{GR value at} \\
        \midrule
        $\delta\phi_{0}$  & [-0.154, 0.363] & -1.34$\sigma$   & $\delta\phi^{\rm PCA}_{0\text{-th}}$ & [-0.159 0.00520] & 2.85$\sigma$    \\
        $\delta\phi_{2.5}$& [-65.1, 57.9]   & -0.0112$\sigma$ & $\delta\phi^{\rm PCA}_{1\text{-st}}$ & [-0.334 0.466]   & -0.480$\sigma$  \\
        $\delta\phi_{3}$  & [-41.2, 50.9]   & -0.0662$\sigma$ & $\delta\phi^{\rm PCA}_{2\text{-nd}}$ & [-2.80 5.06]     & -0.889$\sigma$  \\
        $\delta\phi_{4}$  & [-9.15, 6.77]   & 0.190$\sigma$   & $\delta\phi^{\rm PCA}_{3\text{-rd}}$ & [-71.3 83.0]     & -0.0181$\sigma$ \\
        \bottomrule
    \end{tabular}
    \caption{\textbf{3-$\sigma$ uncertain ranges and distance between GR values and maximum likelihood values of combined results for original dispersion parameters and PCA parameters.}
    The dominant PCA parameters can be better measured and constrained, thus having smaller error bars. The operation of PCA will make the behavior of deviating from zero in posteriors more obvious by transforming posterior samples into the set of new bases. 
    }
    \label{tab_results}
\end{table}

\section{Summmary} \label{sec_summary}

GW observations have opened a new era of exploring the nature of the universe. 
Detected GW events have been leading paradigm-shifting in research of astrophysics, cosmology, and gravity. 
As the launch of the fourth observing run of LVK, it is expected that more GW events will be detected and more amazing discoveries may be revealed. 
GW observations provide a new approach to test gravity theories. Testing gravity with GW has begun to blossom since the first GW event detection in 2015. Using detected events, LVK has performed exhaustive tests of GR \citep{Collaboration2021f,Collaboration2020,Collaboration2021g}. Except the theories agnostic tests performed by LVK, theory-specific tests are also performed by many independent research groups, such as \citep{Perkins2021,Wang2021d,Haegel2023,Gong2022,Wu2022,Ramos2019,Jana2019}. 

In this work, we consider the test of GW dispersion relation which is conventionally performed by LVK. But in tests of LVK, to avoid the correlations among deformation parameters to yield uninformative posteriors, only one deformation parameter is varied at a time. 
This is equivalent to considering the prior of $\delta$ function for the fixed parameters.
If considering the most agnostic prior, all deformation parameters need to be estimated simultaneously.
Previous works \citep{Ohme2013,Pai2012,Saleem2021,Datta2022,Datta2023,Shoom2023} show that the correlations can be reduced by the method of PCA. Using this method, the authors considered the parameterized test of PN structure where multiple deformation parameters are allowed to vary simultaneously, but still obtain informative posteriors meanwhile.
We extend this method to the test of GW dispersion relation in this work.

The most difference of our analyses comparing with the tests of LVK is that the multiple dispersion parameters are varied simultaneously in Bayesian parameter estimation. And then we transform the obtained posterior samples into a set of new bases constructed by the PCA. The new PCA parameters have less correlations, and dominant PCA parameters can be better measured and constrained, thus are expected to be more sensitive to potential deviations from GR.
We consider 10 farthest events selected by criteria of FAR $< 10^{-3}\ \text{y}^{-1}$, SNR $>12$, and $p_{\text{astro}}>0.99$. We consider 4 different GR waveform and add the modifications of dispersion effect into them as the template in Bayesian parameter estimation.  
The results we obtain are consistent with GR within 3-$\sigma$ uncertainty in all cases.
However, we observe that the maximum likelihood value of the first PCA parameter has a relatively large departure from the GR value. This on the one hand shows the dominant PCA parameter is more sensitive to deviations from zero in posteriors, on the other hand hints that the demand of waveform accuracy is higher for using PCA. 
We also find that minor differences in posteriors of original parameters can lead to obvious difference in results after the operation of PCA, which from another aspect indicates multi-parameter tests with PCA require more accurate waveform models.

\begin{acknowledgements}
W.Z. is supported by the National Key R\&D Program of China (Grant No. 2021YFC2203102 and 2022YFC2204602), Strategic Priority Research Program of the Chinese Academy of Science (Grant No. XDB0550300), the National Natural Science Foundation of China (Grant No. 12325301 and 12273035), the Fundamental Research Funds for the Central Universities (Grant No. WK2030000036 and WK3440000004), the Science Research Grants from the China Manned Space Project (Grant No.CMS-CSST-2021-B01), the 111 Project for "Observational and Theoretical Research on Dark Matter and Dark Energy" (Grant No. B23042).
R.N. is supported in part by the National Key Research and Development Program of China Grant No.2022YFC2807303.

This research has made use of data or software obtained from the Gravitational Wave Open Science Center (gwosc.org), a service of the LIGO Scientific Collaboration, the Virgo Collaboration, and KAGRA. This material is based upon work supported by NSF's LIGO Laboratory which is a major facility fully funded by the National Science Foundation, as well as the Science and Technology Facilities Council (STFC) of the United Kingdom, the Max-Planck-Society (MPS), and the State of Niedersachsen/Germany for support of the construction of Advanced LIGO and construction and operation of the GEO600 detector. Additional support for Advanced LIGO was provided by the Australian Research Council. Virgo is funded, through the European Gravitational Observatory (EGO), by the French Centre National de Recherche Scientifique (CNRS), the Italian Istituto Nazionale di Fisica Nucleare (INFN) and the Dutch Nikhef, with contributions by institutions from Belgium, Germany, Greece, Hungary, Ireland, Japan, Monaco, Poland, Portugal, Spain. KAGRA is supported by Ministry of Education, Culture, Sports, Science and Technology (MEXT), Japan Society for the Promotion of Science (JSPS) in Japan; National Research Foundation (NRF) and Ministry of Science and ICT (MSIT) in Korea; Academia Sinica (AS) and National Science and Technology Council (NSTC) in Taiwan.

The numerical calculations in this paper have been done on the supercomputing system in the Supercomputing Center of University of Science and Technology of China.
Data analyses and results visualization in this work made use of  \texttt{Bilby} \citep{Ashton2019}, \texttt{Pymultinest} \citep{Buchner2014,Feroz2008,Feroz2007}, \texttt{LALSuite} \citep{lalsuite}, \texttt{PESummary} \citep{Hoy2021a}, \texttt{NumPy} \citep{Harris2020, Walt2011}, \texttt{Scipy} \citep{Virtanen2020}, and \texttt{matplotlib} \citep{Hunter2007}.

\end{acknowledgements}

\bibliographystyle{raa}
\bibliography{ref}

\begin{thebibliography}{84}
\providecommand\natexlab[1]{#1}
\providecommand\JournalTitle[1]{#1}

\bibitem[Abbott {et~al.}(2020{\natexlab{a}})]{Abbott2020c}
Abbott, B.~P., Abbott, R., Abbott, T.~D., {et~al.} 2020{\natexlab{a}}, Classical and Quantum Gravity, 37, 055002

\bibitem[Abbott {et~al.}(2016{\natexlab{a}})]{Abbott2016}
Abbott, B., Abbott, R., Abbott, T., {et~al.} 2016{\natexlab{a}}, Physical Review Letters, 116, 061102

\bibitem[Abbott {et~al.}(2016{\natexlab{b}})]{Abbott2016a}
Abbott, B., Abbott, R., Abbott, T., {et~al.} 2016{\natexlab{b}}, Physical Review Letters, 116, 221101

\bibitem[Abbott {et~al.}(2019{\natexlab{a}})]{Abbott2019}
Abbott, B., Abbott, R., Abbott, T., {et~al.} 2019{\natexlab{a}}, Physical Review X, 9, 031040

\bibitem[Abbott {et~al.}(2019{\natexlab{b}})]{Abbott2019c}
Abbott, B., Abbott, R., Abbott, T., {et~al.} 2019{\natexlab{b}}, Physical Review Letters, 123, 011102

\bibitem[Abbott {et~al.}(2019{\natexlab{c}})]{Abbott2019b}
Abbott, B., Abbott, R., Abbott, T., {et~al.} 2019{\natexlab{c}}, Physical Review D, 100, 104036

\bibitem[Abbott {et~al.}(2020{\natexlab{b}})]{Abbott2020}
Abbott, R., Abbott, T.~D., Abraham, S., {et~al.} 2020{\natexlab{b}}, arXiv:2010.14527

\bibitem[Adelberger(2001)]{Adelberger2001}
Adelberger, E.~G. 2001, Classical and Quantum Gravity, 18, 2397

\bibitem[Agathos {et~al.}(2014)]{Agathos2014}
Agathos, M., Pozzo, W.~D., Li, T., {et~al.} 2014, Physical Review D, 89, 082001

\bibitem[Aghanim {et~al.}(2020)]{Aghanim2020}
Aghanim, N., Akrami, Y., Ashdown, M., {et~al.} 2020, Astronomy and Astrophysics, 641, A6

\bibitem[Amelino-Camelia(2002)]{AmelinoCamelia2002}
Amelino-Camelia, G. 2002, Nature, 418, 34

\bibitem[Ashton \& Khan(2019)]{Ashton2019a}
Ashton, G., \& Khan, S. 2019, Phys. Rev. D 101, 064037 (2020), 101, 064037

\bibitem[Ashton {et~al.}(2019)]{Ashton2019}
Ashton, G., Hübner, M., Lasky, P.~D., {et~al.} 2019, The Astrophysical Journal Supplement Series, 241, 27

\bibitem[Belgacem {et~al.}(2018)]{Belgacem2018}
Belgacem, E., Dirian, Y., Foffa, S., \& Maggiore, M. 2018, Physical Review D, 97, 104066

\bibitem[Berti {et~al.}(2015)]{Berti2015}
Berti, E., Barausse, E., Cardoso, V., {et~al.} 2015, arXiv:1501.07274

\bibitem[Buchner {et~al.}(2014)]{Buchner2014}
Buchner, J., Georgakakis, A., Nandra, K., {et~al.} 2014, Astronomy \&amp; Astrophysics, 564, A125

\bibitem[Calcagni(2010)]{Calcagni2010}
Calcagni, G. 2010, Physical Review Letters, 104, 251301

\bibitem[Collaboration {et~al.}(2020)]{Collaboration2020}
Collaboration, T. L.~S., the Virgo~Collaboration, Abbott, R., {et~al.} 2020, arXiv:2010.14529

\bibitem[Collaboration {et~al.}(2021{\natexlab{a}})]{Collaboration2021e}
Collaboration, T. L.~S., the Virgo~Collaboration, Abbott, R., {et~al.} 2021{\natexlab{a}}, arXiv:2108.01045

\bibitem[Collaboration {et~al.}(2021{\natexlab{b}})]{Collaboration2021f}
Collaboration, T. L.~S., the Virgo~Collaboration, the KAGRA~Collaboration, {et~al.} 2021{\natexlab{b}}, arXiv:2111.03606

\bibitem[Collaboration {et~al.}(2021{\natexlab{c}})]{Collaboration2021g}
Collaboration, T. L.~S., the Virgo~Collaboration, the KAGRA~Collaboration, {et~al.} 2021{\natexlab{c}}, arXiv:2112.06861

\bibitem[Cornish \& Littenberg(2015)]{Cornish2015}
Cornish, N.~J., \& Littenberg, T.~B. 2015, Classical and Quantum Gravity, 32, 135012

\bibitem[Datta(2023)]{Datta2023}
Datta, S. 2023, arXiv:2303.04399

\bibitem[Datta {et~al.}(2020)]{Datta2020}
Datta, S., Gupta, A., Kastha, S., Arun, K.~G., \& Sathyaprakash, B.~S. 2020, arXiv:2006.12137

\bibitem[Datta {et~al.}(2022)]{Datta2022}
Datta, S., Saleem, M., Arun, K.~G., \& Sathyaprakash, B.~S. 2022, arXiv:2208.07757

\bibitem[Debono \& Smoot(2016)]{Debono2016}
Debono, I., \& Smoot, G. 2016, Universe, 2, 23

\bibitem[DeWitt(1967)]{DeWitt1967}
DeWitt, B.~S. 1967, Physical Review, 160, 1113

\bibitem[Feroz \& Hobson(2007)]{Feroz2007}
Feroz, F., \& Hobson, M.~P. 2007, Mon. Not. Roy. Astron. Soc., 384, 2, 449-463 (2008), 384, 449

\bibitem[Feroz {et~al.}(2008)]{Feroz2008}
Feroz, F., Hobson, M.~P., \& Bridges, M. 2008, Mon. Not. Roy. Astron. Soc. 398: 1601-1614,2009, 398, 1601

\bibitem[Foreman-Mackey {et~al.}(2013)]{ForemanMackey2013}
Foreman-Mackey, D., Hogg, D.~W., Lang, D., \& Goodman, J. 2013, Publications of the Astronomical Society of the Pacific, 125, 306

\bibitem[Frieman {et~al.}(2008)]{Frieman2008}
Frieman, J.~A., Turner, M.~S., \& Huterer, D. 2008, Annual Review of Astronomy and Astrophysics, 46, 385

\bibitem[Ghosh {et~al.}(2016)]{Ghosh2016}
Ghosh, A., Ghosh, A., Johnson-McDaniel, N.~K., {et~al.} 2016, Physical Review D, 94, 021101

\bibitem[Ghosh {et~al.}(2017)]{Ghosh2017}
Ghosh, A., Johnson-McDaniel, N.~K., Ghosh, A., {et~al.} 2017, Classical and Quantum Gravity, 35, 014002

\bibitem[Gong {et~al.}(2022)]{Gong2022}
Gong, C., Zhu, T., Niu, R., {et~al.} 2022, Physical Review D, 105, 044034

\bibitem[Gupta {et~al.}(2020)]{Gupta2020}
Gupta, A., Datta, S., Kastha, S., {et~al.} 2020, Phys. Rev. Lett. 125, 201101 (2020), 125, 201101

\bibitem[Haegel {et~al.}(2023)]{Haegel2023}
Haegel, L., O'Neal-Ault, K., Bailey, Q.~G., {et~al.} 2023, Physical Review D, 107, 064031

\bibitem[Harris {et~al.}(2020)]{Harris2020}
Harris, C.~R., Millman, K.~J., van~der Walt, S.~J., {et~al.} 2020, Nature, 585, 357

\bibitem[Hastings(1970)]{Hastings1970}
Hastings, W.~K. 1970, Biometrika, 57, 97

\bibitem[Ho{\v{r}}ava(2009)]{Horava2009}
Ho{\v{r}}ava, P. 2009, Physical Review D, 79, 084008

\bibitem[Hoy \& Raymond(2021)]{Hoy2021a}
Hoy, C., \& Raymond, V. 2021, {SoftwareX}, 15, 100765

\bibitem[Hoyle {et~al.}(2001)]{Hoyle2001}
Hoyle, C.~D., Schmidt, U., Heckel, B.~R., {et~al.} 2001, Physical Review Letters, 86, 1418

\bibitem[Hunter(2007)]{Hunter2007}
Hunter, J.~D. 2007, Computing in Science {\&} Engineering, 9, 90

\bibitem[Jain \& Khoury(2010)]{Jain2010}
Jain, B., \& Khoury, J. 2010, Annals of Physics, 325, 1479

\bibitem[Jana \& Mohanty(2019)]{Jana2019}
Jana, S., \& Mohanty, S. 2019, Physical Review D, 99, 044056

\bibitem[Kiefer(2007)]{Kiefer2007}
Kiefer, C. 2007, in Approaches to Fundamental Physics (Springer Berlin Heidelberg), 123

\bibitem[Kosteleck{\'{y}} \& Mewes(2016)]{Kostelecky2016}
Kosteleck{\'{y}}, V.~A., \& Mewes, M. 2016, Physics Letters B, 757, 510

\bibitem[Koyama(2016)]{Koyama2016}
Koyama, K. 2016, Reports on Progress in Physics, 79, 046902

\bibitem[Kramer(2017)]{Kramer2017}
Kramer, M. 2017, in The Fourteenth Marcel Grossmann Meeting ({WORLD} {SCIENTIFIC})

\bibitem[Li {et~al.}(2012)]{Li2012}
Li, T. G.~F., Pozzo, W.~D., Vitale, S., {et~al.} 2012, Physical Review D, 85, 082003

\bibitem[{LIGO Scientific Collaboration}(2018)]{lalsuite}
{LIGO Scientific Collaboration}. 2018, {LIGO} {A}lgorithm {L}ibrary - {LALS}uite, free software (GPL)

\bibitem[Manchester(2015)]{Manchester2015}
Manchester, R.~N. 2015, International Journal of Modern Physics D, 24, 1530018

\bibitem[Mirshekari {et~al.}(2012)]{Mirshekari2012}
Mirshekari, S., Yunes, N., \& Will, C.~M. 2012, 85, 024041

\bibitem[Nishizawa(2018)]{Nishizawa2018}
Nishizawa, A. 2018, Physical Review D, 97, 104037

\bibitem[Niu {et~al.}(2024)]{Niu2024}
Niu, R., Ma, Z.-C., Chen, J.-M., Feng, C., \& Zhao, W. 2024, Results in Physics, 107407

\bibitem[Niu {et~al.}(2022)]{Niu2022}
Niu, R., Zhu, T., \& Zhao, W. 2022, Journal of Cosmology and Astroparticle Physics, 2022, 011

\bibitem[Ohme {et~al.}(2013)]{Ohme2013}
Ohme, F., Nielsen, A.~B., Keppel, D., \& Lundgren, A. 2013, Physical Review D, 88, 042002

\bibitem[Okounkova {et~al.}(2021)]{Okounkova2021}
Okounkova, M., Farr, W.~M., Isi, M., \& Stein, L.~C. 2021, Phys. Rev. D 106, 044067 (2022), 106, 044067

\bibitem[Pai \& Arun(2012)]{Pai2012}
Pai, A., \& Arun, K.~G. 2012, Classical and Quantum Gravity, 30, 025011

\bibitem[Perkins {et~al.}(2021)]{Perkins2021}
Perkins, S.~E., Nair, R., Silva, H.~O., \& Yunes, N. 2021, arXiv:2104.11189

\bibitem[Porter {et~al.}(2011)]{Porter2011}
Porter, T.~A., Johnson, R.~P., \& Graham, P.~W. 2011, Annual Review of Astronomy and Astrophysics, 49, 155

\bibitem[Pratten {et~al.}(2020)]{Pratten2020}
Pratten, G., Husa, S., Garc{\'{\i}}a-Quir{\'{o}}s, C., {et~al.} 2020, Physical Review D, 102, 064001

\bibitem[Pratten {et~al.}(2021)]{Pratten2021}
Pratten, G., Garc{\'{\i}}a-Quir{\'{o}}s, C., Colleoni, M., {et~al.} 2021, Physical Review D, 103, 104056

\bibitem[Pürrer(2014)]{Puerrer2014}
Pürrer, M. 2014, Class. Quantum Grav. 31 195010, 2014, 31, 195010

\bibitem[Pürrer(2015)]{Puerrer2015}
Pürrer, M. 2015, Phys. Rev. D 93, 064041 (2016), 93, 064041

\bibitem[Ramos \& Barausse(2019)]{Ramos2019}
Ramos, O., \& Barausse, E. 2019, Physical Review D, 99, 024034

\bibitem[Saleem {et~al.}(2021)]{Saleem2021}
Saleem, M., Datta, S., Arun, K.~G., \& Sathyaprakash, B.~S. 2021, arXiv:2110.10147

\bibitem[Sefiedgar {et~al.}(2011)]{Sefiedgar2011}
Sefiedgar, A., Nozari, K., \& Sepangi, H. 2011, Physics Letters B, 696, 119

\bibitem[Shoom {et~al.}(2023)]{Shoom2023}
Shoom, A.~A., Gupta, P.~K., Krishnan, B., Nielsen, A.~B., \& Capano, C.~D. 2023, General Relativity and Gravitation, 55

\bibitem[Skilling(2004)]{Skilling2004}
Skilling, J. 2004, in {AIP} Conference Proceedings ({AIP})

\bibitem[Skilling(2006)]{Skilling2006}
Skilling, J. 2006, Bayesian Analysis, 1, 833

\bibitem[Stairs(2003)]{Stairs2003}
Stairs, I.~H. 2003, Living Reviews in Relativity, 6

\bibitem[van~der Walt {et~al.}(2011)]{Walt2011}
van~der Walt, S., Colbert, S.~C., \& Varoquaux, G. 2011, Computing in Science {\&} Engineering, 13, 22

\bibitem[Varma {et~al.}(2019)]{Varma2019}
Varma, V., Field, S.~E., Scheel, M.~A., {et~al.} 2019, Phys. Rev. Research 1, 033015 (2019), 1, 033015

\bibitem[Virtanen {et~al.}(2020)]{Virtanen2020}
Virtanen, P., Gommers, R., Oliphant, T.~E., {et~al.} 2020, Nature Methods, 17, 261

\bibitem[Wang {et~al.}(2021{\natexlab{a}})]{Wang2021d}
Wang, Y.-F., Brown, S.~M., Shao, L., \& Zhao, W. 2021{\natexlab{a}}, arXiv:2109.09718

\bibitem[Wang {et~al.}(2022)]{Wang2022b}
Wang, Y.-F., Brown, S.~M., Shao, L., \& Zhao, W. 2022, Physical Review D, 106, 084005

\bibitem[Wang {et~al.}(2021{\natexlab{b}})]{Wang2021g}
Wang, Y.-F., Niu, R., Zhu, T., \& Zhao, W. 2021{\natexlab{b}}, The Astrophysical Journal, 908, 58

\bibitem[Wex(2014)]{Wex2014}
Wex, N. 2014, arXiv:1402.5594

\bibitem[Will(1998)]{Will1998}
Will, C.~M. 1998, Physical Review D, 57, 2061

\bibitem[Will(2014)]{Will2014}
Will, C.~M. 2014, Living Reviews in Relativity, 17

\bibitem[Wu {et~al.}(2022)]{Wu2022}
Wu, Q., Zhu, T., Niu, R., Zhao, W., \& Wang, A. 2022, Physical Review D, 105, 024035

\bibitem[Yunes \& Pretorius(2009)]{Yunes2009}
Yunes, N., \& Pretorius, F. 2009, Physical Review D, 80, 122003

\bibitem[Yunes {et~al.}(2016)]{Yunes2016}
Yunes, N., Yagi, K., \& Pretorius, F. 2016, Physical Review D, 94, 084002

\bibitem[Zhao {et~al.}(2020)]{Zhao2020a}
Zhao, W., Zhu, T., Qiao, J., \& Wang, A. 2020, Physical Review D, 101, 024002

\end{thebibliography}

\end{document}